\documentclass[epj]{svjour}
\usepackage{latexsym}
\usepackage{graphics}
\usepackage{graphicx}

\newcommand{\tr}{{\textrm {tr}}}

\begin{document}
\title{Dimension 2 condensates and Polyakov Chiral Quark Models}
\author{\underline{E. Meg{\'\i}as}
\thanks{\emph{Speaker at QNP06 Madrid (Spain), 5-10 June 2006. 
Work supported by the Spanish DGI and FEDER funds grant
FIS2005-00810, Junta de Andaluc{\'\i}a grant FQM-225, 
and EURIDICE Grant No. HPRN-CT-2002-00311.} }
\and E. Ruiz Arriola 
\and L.L. Salcedo 
}                     
\institute{
Departamento de F{\'\i}sica At\'omica, Molecular y
Nuclear, Universidad de Granada, E-18071 Granada, Spain}
\date{\today}
%
\abstract{We address a possible relation between the expectation value
of the Polyakov loop in pure gluodynamics and full QCD based on
Polyakov Chiral Quark Models where constituent quarks and the Polyakov
loop are coupled in a minimal way. To this end we use a center
symmetry breaking Gaussian model for the Polyakov loop distribution
which accurately reproduces gluodynamics data above the phase
transition in terms of dimension 2 gluon condensate. The role played
by the quantum and local nature of the Polyakov loop is emphasized.
} 
%

\maketitle



The phase transition of QCD matter at finite
temperature from hadrons to a quark-gluon plasma was established long
ago~\cite{Kogut:1982rt,Fukugita:1986rr,Karsch:1994hm} (for a review
see e.g. ~\cite{Karsch:1998hr}). 
The precise definition of the phase transition requires a proper
identification of the relevant order parameters. In the heavy quark
limit, the Polyakov loop vacuum expectation value signals the breaking
of the center symmetry corresponding to the deconfinement phase
transition when changing from zero to one~\cite{Pisarski:2002ji}. In
the limit of light quarks the chiral condensate determines the
restoration of chiral symmetry when the chiral condensate vanishes
above the critical temperature. In the real QCD case with dynamical
massive quarks both chiral and center symmetries are explicitly
broken, and neither the Polyakov loop nor the chiral condensate are
truly order parameters, although a rather sharp crossover is expected
across the phase transition for these quantities~\cite{Karsch:1998hr}.

Polyakov-Chiral Quark Models allow to study the interplay between
chiral symmetry restoration and center symmetry breaking
\cite{Mocsy:2003qw} in a quantitative
manner~\cite{Meisinger:1995ih,Fukushima:2003fw,Megias:2004kc,Megias:2004gy,Megias:2004hj,Ratti:2005jh,Ghosh:2006qh}. At
zero temperature the constituent quark mass, $M$, is dynamically
generated via the spontaneous breaking of chiral symmetry, inducing a
exponentially small, $\sim e^{-M/T}$, breaking of the center symmetry
at low
temperatures~\cite{Megias:2004kc,Megias:2004gy,Megias:2004hj}. This
provides the rationale for keeping the Polyakov loop as an order
parameter also in the unquenched case. However, although the coupling
of the Polyakov loop to quarks is rather unique, details regarding the
postulated purely gluonic action
differ~\cite{Meisinger:1995ih,Fukushima:2003fw,Megias:2004kc,Megias:2004gy,Megias:2004hj,Ratti:2005jh,Ghosh:2006qh}. This
additional information is beyond the chiral quark model capabilities
and must always be postulated. In this regard, it seems natural to
constrain the gluonic action to reproduce pure gluodynamics
results. In the present work we address a possible relation between
the expectation value of the Polyakov loop in pure gluodynamics and
full QCD based on the coupling to quarks generated by the fermion
determinant within a Polyakov NJL model (PNJL).

One of the problems one must also face when comparing models with lattice data
has to do with the difficult but necessary renormalization of the Polyakov
loop. Indeed, after renormalization the Polyakov loop falls outside the
unitary group~\cite{Gava:1981qd,Zantow:2003uh}.  The spontaneous breaking of
the center symmetry above some critical temperature occurs already at the
level of pure gluodynamics and is interpreted as the signal of
deconfinement~\cite{Kaczmarek:2002mc}. Full dynamical QCD lattice simulations
account, in addition, for an explicit center symmetry breaking at low
temperatures due to the presence of fermions~\cite{Kaczmarek:2005ui}. In a
recent paper~\cite{Megias:2005ve} we have shown that the behaviour of the
Polyakov loop above the deconfinement phase transition can be naturally
accommodated by a dimension-2 condensate, instead of the more conventional
perturbative calculations. The quality of our fits leaves little doubt on the
veracity of the description of the lattice data.  To our knowledge this is the
first time that power corrections in temperature have been advocated to
describe QCD right above the deconfinement phase transition. This is in sharp
contrast to the common believe that logarithmic thermal corrections should be
considered instead yielding to rather unnatural, in fact inaccurate, fits to
the existing lattice results.

Following our study on the role of dimension-2 condensate in the
center symmetry breaking as a guideline~\cite{Megias:2005ve} we make
here a Gaussian ansatz for the purely gluonic contribution to the
Polyakov loop distribution which reproduces remarkably well the
lattice data for pure gluodynamics~\cite{Kaczmarek:2002mc}.  Then,
using the PNJL model~\cite{Meisinger:1995ih,Fukushima:2003fw,Megias:2004kc,Megias:2004gy,Megias:2004hj,Ratti:2005jh,Ghosh:2006qh}
we may {\it predict} the value of the Polyakov loop in the {\it
unquenched} case and compare with full dynamical calculations. In our
view this is the best possible scenario for a Polyakov-Chiral quark
model to work.


Let us explain how the dimension-2 gluon condensate can generate a power like
behaviour of the Polyakov loop at temperatures slightly above the
deconfinement phase transition in gluodynamics. In the Polyakov gauge the
(expectation value of the) Polyakov loop is defined as
\begin{equation}
L = \left\langle \frac{1}{N_c}\tr_c \, e^{ig A_0(\vec{x})/T } \right\rangle
\,,
\label{eq:10}
\end{equation}
where $\langle~\rangle$ denotes vacuum expectation value and $\tr_c$
is the (fundamental) color trace.
$A_0$ is the gluon field in the (Euclidean) time direction, $A_0 = \sum T_a
A_{0,a}$, $T_a$ being the Hermitian generators of the SU$(N_c)$ Lie
algebra in the fundamental representation, with the standard
normalization $\tr_c(T_aT_b)=\delta_{ab}/2$. Note that in this gauge we
have the invariance under the large gauge transformation $ g A_0\to g
A_0 + 2 \pi T$.


If we expand the Polyakov loop at high temperatures we get the
cumulant  expansion
\begin{eqnarray}
L &=& \sum_{k=0}^\infty \frac1{(2k)!}\left( \frac{-g^2}{T^2} \right)^k
\langle A_{0,a_1} \dots A_{0,a_{2k}} \rangle \nonumber \\ &\times &
\frac1{N_c} \tr_c ( T_{a_1} \dots T_{a_{2k}} )
\end{eqnarray} 
where odd terms have been discarded on the assumption that global
colour symmetry breaking does not occur. Averaging over the $N_c^2-1$
colour gluonic degrees of freedom we get
\begin{eqnarray}
\langle A_{0,a_1} \dots A_{0,a_{2k}} \rangle = \frac{ \langle
(A_{0,a})^{2k}\rangle (N_c^2-3) !!}{(N_c^2+2 k-3)!!} \delta_{a_1,\dots
, a_{2k}} 
\end{eqnarray} 
where $\delta_{a_1,\dots , a_{2k}}$ is the fully symmetrized product of
the Kronecker delta symbols $\delta_{ab}$. The SU$(N_c)$ Casimir
invariant is  $C_2 = \sum_a T_a T_a = (N_c^2-1)/(2 N_c)  {\bf 1}_{N_c}$.   
In the large $N_c$ limit only the contractions of adjacent indices
(including cyclic permutations) survive.  For instance, if we twist
two adjacently contracted generators we get $ T_a T_b T_a = - T_b /2
N_c $ so that $ (T_a T_b)^2 $ is suppressed by two powers of $N_c $ as
compared to $ T_a^2 T_b^2 $. Then, neglecting these terms we obtain
\begin{eqnarray} 
L  = \sum_{k=0}^\infty \frac{1}{(2 k ) ! } \left(\frac{-g^2}{T^2 } \right)^k
\frac{\langle (A_{0,a})^{2k}\rangle}{ (2
N_c)^k} \,.
\end{eqnarray} 
Now, we expect $ \langle (A_{0,a})^{2k}\rangle $ to scale as $ N_c^{2
k} $, in which case and for large $T$ the limit $N_c \to \infty $ with
$g^2 N_c$ fixed is finite and nontrivial, at least in the absence of
radiative corrections. Finally, if we assume vacuum saturation of
condensates we can reduce further the expression since 
\begin{eqnarray}
\langle A_0^{2k} \rangle = (2 k-1) !! \langle A_0^{2} \rangle^k + {\rm
n.v.c.}
\end{eqnarray} 
where n.v.c stands for non vacuum connected terms.  After summing up
 the series we get the result
\begin{equation}
 L  = \exp\left[-\frac{g^2\langle A_{0,a}^2\rangle}{4N_cT^2}
 \right] \,.
\label{eq:L_exp}
\end{equation}
The exponentiation means that temperature is not necessarily very high
since we summed up all the series and, actually, for small $T$ yields
$L \to 0$. Of course, as one approaches the true critical temperature
one expects non analytical behaviour not encoded in the high
temperature expansion~\footnote{Nonetheless, although
Eq.~(\ref{eq:L_exp}) does not predict a phase transition one could
still make an estimate of the critical temperature related to the
point where $L$ presents no curvature, i.e. the
inflexion point which yields, $T_c^2 = g^2 \langle A_0^2 \rangle /6
N_c $. On the other hand we expect the condensate to scale with $N_c $
and the number of Euclidean dimensions as $ g^2 \langle A_\mu^2
\rangle_\mu \sim 4 \pi \alpha (\mu) (N_c^2 - 1) D \Lambda_{\rm
QCD}^2$. This shows that the large $N_c$ limit of $L$ is well defined
for {\it fixed} T. Taking $\Lambda_{\rm
QCD} = 240 \,{\rm MeV} $ and the scale
$\mu=2 \,{\rm GeV} $ with the conversion factor $\mu = 2\pi T$ and $N_c
= 3 $ we get $g^2 \langle A_\mu^2 \rangle = (2.2 \, {\rm GeV})^2 $ in
rough agreement with the determinations from lattice data. Then $T_c^2
\sim 4 \pi \alpha ( 2 \pi T_c) (N_c^2 -1) \Lambda_{\rm
QCD}^2/ 6 N_c $ yielding the
estimate $T_c=1.1 \Lambda_{\rm QCD} $.}. On top of
Eq.~(\ref{eq:L_exp}) one must take radiative corrections into account
yielding, in fact, $L > 1 $~\cite{Gava:1981qd,Zantow:2003uh} because
of renormalization effects at high temperatures. This is equivalent to
multiply Eq.~(\ref{eq:L_exp}) by a slowly varying temperature
dependent factor~\cite{Megias:2005ve}.

%


The previous behaviour suggests a Gaussian ansatz for the Polyakov
loop distribution, i.e. the probability of having a configuration with
a given value of $\Omega$. Unfortunately, not much is known about this
object in QCD (see however \cite{Megias:2004hj}). In the Polyakov
gauge, where $A_0 $ is both static and diagonal we can write the piece
of the vacuum wave functional depending on the temporal gluon fields
as,
\begin{eqnarray}
\rho_0 ( A_3 , A_8) = C e^{-( A_3^2 + A_8^2 )/ a^2}  
\end{eqnarray} 
where we assume that the gluon variables take any real values on the
real axis. The normalization constant $C$ and the width of the
distribution $a$ can be determined from the definition of the
dimension 2 condensate
\begin{eqnarray}
\langle  A_3^2 + A_8^2 \rangle_0 = a^2  \,. 
\end{eqnarray} 
Although the previous Gaussian reproduces the result of the cumulant expansion
of the Polyakov loop, it does not incorporate the known gauge invariance which
corresponds to a periodicity condition on the gluon field distribution and the
integration measure.

Once the Polyakov loop distribution has been determined for pure
gluodynamics we set out to describe its expectation value when
dynamical quarks are included. To do so we use the PNJL model.
For a constant value of the Polyakov loop, $\Omega$, and the scalar
field $S=M$ (which we identify with the constituent quark mass) the
quark effective action is given by 
\begin{eqnarray}
\frac{  T }{V} \Gamma_Q &=& \frac{1}{4 G_S} {\rm Tr}_f(M
- \hat{M}_0)^2- 2 N_f \int \frac{d^3 k}{(2\pi)^3}
\Big(
 N_c \epsilon_k
\nonumber \\ && 
+ T \, \tr_c 
\log\left[ 1 + e^{-\epsilon_k/T} \Omega \right] + {\rm c.c.} 
\Big)
\label{eq:gammaQ1}
\end{eqnarray} 
where we have only retained the vacuum contribution, so there is no
contribution of meson fields. $V$ is three dimensional
volume and $\epsilon_k = +\sqrt{k^2 + M^2 } $ is the energy of a
constituent quark with mass $M$. We define the Polyakov-loop averaged
action
\begin{eqnarray}
e^{-\Gamma_Q (M,T)} = \int d \Omega \, \rho_0(A_3,A_8)
\, e^{-\Gamma_Q (M,\Omega,T)} \,.
\end{eqnarray} 
The value of $M$ is determined by minimization of $\Gamma_Q(M,T)$ with respect
to $M$, $ \partial \Gamma_Q (M,T) / \partial M=0 $, which is the gap equation
and determines $M$ at a given temperature $T$, denoted as $M^* = M(T) $. This
procedure allows to compute the integration in meson fields at the mean field
level. In addition, the relation between the (single flavour) chiral quark
condensate, $\langle \bar q q \rangle$, and the constituent quark mass, reads
\begin{eqnarray} 
2 G_S \langle \bar q q \rangle^* = - (M^* - m_q) \,.
\label{eq:gap*}
\end{eqnarray} 
Any observable is obtained by using the temperature dependent mass
$M^*$. For instance the expectation value of the Polyakov loop is
computed as
\begin{eqnarray}
\langle \Omega \rangle = \frac{\int d \Omega \, \rho_0(A_3,A_8)  
\, e^{-\Gamma_Q (M^*,\Omega,T)} \,\Omega }{\int d \Omega \,
\rho_0(A_3,A_8) \,e^{-\Gamma_Q (M^*,\Omega,T)}} \,.
\end{eqnarray} 
The integral in $d\Omega$ in the case $N_c=3$ and in the Polyakov
gauge can be computed numerically.  At the mean field level the Polyakov
distribution becomes a delta function, but this picture is
obviously only valid at high enough temperature, where quantum effects
are negligible. At low temperatures the mean field approximation can
lead to wrong results. For example, for the expectation value of the
Polyakov loop in the adjoint representation one obtains $-1/(N_c^2-1)$,
instead of zero, as has been observed in lattice
data~\cite{Dumitru:2003hp}. The color group integration solves this
problem~\cite{Megias:2004hj}.  Finally, we have taken into account here the 
local effects in the Polyakov loop by consider a confinement
correlation volume $V_\sigma=8\pi T^3/\sigma^3$, which follows from
the simplest correlation of two Polyakov loops
\begin{equation}
\frac{1}{T}\int d^3x \int d\Omega \, \tr_c \Omega(\vec{x}) \,\tr_c
\Omega^\dagger(\vec{y}) = \frac{1}{T} \int d^3x \, e^{-\sigma
  |\vec{x}-\vec{y}|/T} \,. 
\end{equation}
Higher correlation functions are expected to have a smaller
contribution~\cite{Megias:2004hj}.

\begin{figure}[tb]
\vspace{5mm}
\begin{center}
\includegraphics[angle=0,width=0.44\textwidth]{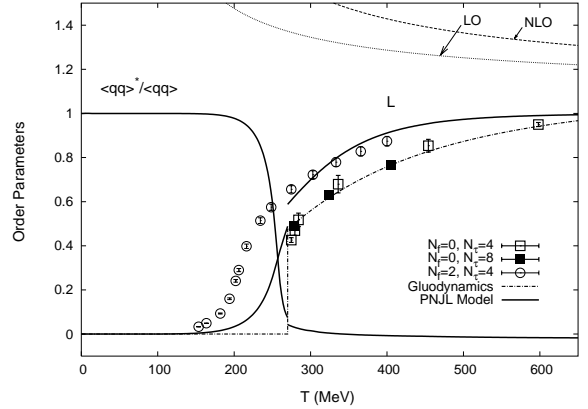}
\end{center}
\vspace{1mm}
\caption{
  Temperature dependence of chiral condensate $\langle \bar q q \rangle^*$ and
  Polyakov loop expectation value $L= \langle \tr_c \Omega \rangle/N_c$ in the
  2-flavor PNJL model.  We use the Gaussian distribution
  reproducing the Polyakov loop in the case of pure gluodynamics. Lattice data
  from \cite{Kaczmarek:2002mc} for gluodynamics ($N_f=0$) and
  \cite{Kaczmarek:2005ui} for $N_f=2$ full QCD. Perturbative LO and NLO
  results for the Polyakov loop with $N_f=2$ are shown for comparison.  }
\label{fig:loop}
\end{figure}

The result for the chiral condensate and Polyakov loop expectation value is
presented in Fig.~\ref{fig:loop}.  We consider $\sqrt{\sigma}= 425\, {\rm
  MeV}$ and $m_u = m_d \equiv m_q = 5.5 \, {\rm MeV} $. As we see the agreement
is reasonable although the quark model predicts a shift in the critical
temperature of about $50 \, {\rm MeV}$. This discrepancy is compatible with
the model uncertainties discussed in our previous work~\cite{Megias:2004hj}
where the explicit center symmetry breaking of the pure gluonic action was
disregarded. It is also compatible with the large rescaling advocated in
\cite{Ratti:2005jh}.  Basically, the shift in the critical temperature
reflects our lack of knowledge on the gluon action at intermediate
temperatures, in particular, the feedback of quarks on the gluons. It would be
most useful to develop a theoretical framework where this feedback is taken
into account, as it is done, e.g. in the mean field approximation
\cite{Ratti:2005jh}, while retaining the fundamentally quantum and local
nature of the Polyakov loop \cite{Megias:2004hj}.

%


\begin{thebibliography}{34}
\expandafter\ifx\csname natexlab\endcsname\relax\def\natexlab#1{#1}\fi
\expandafter\ifx\csname bibnamefont\endcsname\relax
  \def\bibnamefont#1{#1}\fi
\expandafter\ifx\csname bibfnamefont\endcsname\relax
  \def\bibfnamefont#1{#1}\fi
\expandafter\ifx\csname citenamefont\endcsname\relax
  \def\citenamefont#1{#1}\fi
\expandafter\ifx\csname url\endcsname\relax
  \def\url#1{\texttt{#1}}\fi
\expandafter\ifx\csname urlprefix\endcsname\relax\def\urlprefix{URL }\fi
\providecommand{\bibinfo}[2]{#2}
\providecommand{\eprint}[2][]{\url{#2}}

\bibitem{Kogut:1982rt}
\bibinfo{author}{\bibfnamefont{J.~B.} \bibnamefont{Kogut}}
  \bibnamefont{et~al.}, \bibinfo{journal}{Phys. Rev. Lett.}
  \textbf{\bibinfo{volume}{50}}, \bibinfo{pages}{393} (\bibinfo{year}{1983}).

\bibitem{Fukugita:1986rr}
\bibinfo{author}{\bibfnamefont{M.}~\bibnamefont{Fukugita}} \bibnamefont{and}
  \bibinfo{author}{\bibfnamefont{A.}~\bibnamefont{Ukawa}},
  \bibinfo{journal}{Phys. Rev. Lett.} \textbf{\bibinfo{volume}{57}},
  \bibinfo{pages}{503} (\bibinfo{year}{1986}).

\bibitem{Karsch:1994hm}
\bibinfo{author}{\bibfnamefont{F.}~\bibnamefont{Karsch}} \bibnamefont{and}
  \bibinfo{author}{\bibfnamefont{E.}~\bibnamefont{Laermann}},
  \bibinfo{journal}{Phys. Rev.} \textbf{\bibinfo{volume}{D50}},
  \bibinfo{pages}{6954} (\bibinfo{year}{1994}), \eprint{hep-lat/9406008}.

\bibitem{Karsch:1998hr}
\bibinfo{author}{\bibfnamefont{F.}~\bibnamefont{Karsch}}
  (\bibinfo{year}{1998}), \bibinfo{journal}{PoS}
 \textbf{\bibinfo{volume}{Corfu98}}, \bibinfo{pages}{008}
  (\bibinfo{year}{1999}).

\bibitem{Pisarski:2002ji}
\bibinfo{author}{\bibfnamefont{R.~D.} \bibnamefont{Pisarski}}
  (\bibinfo{year}{2002}), \eprint{hep-ph/0203271}.

\bibitem{Mocsy:2003qw}
   A.~Mocsy, F.~Sannino and K.~Tuominen,
   Phys.\ Rev.\ Lett.\  {\bf 92}, 182302 (2004),
   \eprint{hep-ph/0308135}.

\bibitem{Meisinger:1995ih}
\bibinfo{author}{\bibfnamefont{P.~N.} \bibnamefont{Meisinger}}
  \bibnamefont{and} \bibinfo{author}{\bibfnamefont{M.~C.}
  \bibnamefont{Ogilvie}}, \bibinfo{journal}{Phys. Lett.}
  \textbf{\bibinfo{volume}{B379}}, \bibinfo{pages}{163}
  (\bibinfo{year}{1996}{\natexlab{b}}), \eprint{hep-lat/9512011}.

\bibitem{Fukushima:2003fw}
\bibinfo{author}{\bibfnamefont{K.}~\bibnamefont{Fukushima}},
  \bibinfo{journal}{Phys. Lett.} \textbf{\bibinfo{volume}{B591}},
  \bibinfo{pages}{277} (\bibinfo{year}{2004}), \eprint{hep-ph/0310121}.

\bibitem{Megias:2004kc}
\bibinfo{author}{\bibfnamefont{E.}~\bibnamefont{Meg{\'\i}as}},
  \bibinfo{author}{\bibfnamefont{E.~R.} \bibnamefont{Arriola}},
  \bibnamefont{and} \bibinfo{author}{\bibfnamefont{L.~L.}
  \bibnamefont{Salcedo}} (\bibinfo{year}{2004}{\natexlab{b}}),
  \bibinfo{note}{mini-Workshop on Quark Dynamics: Bled 2004, Bled, Slovenia,
  12-19 Jul 2004}, \eprint{hep-ph/0410053}.

\bibitem{Megias:2004gy}
E.~Meg{\'\i}as, E.~Ruiz~Arriola, and L.~L. Salcedo, \emph{AIP Conf. Proc.}
  \textbf{756}, 436--438 (2005), \eprint{hep-ph/0411293}.

\bibitem{Megias:2004hj}
E.~Meg{\'\i}as, E.~Ruiz~Arriola, and L.~L. Salcedo, \emph{Phys. Rev.} \textbf{D74},
  065005 (2006), \eprint{hep-ph/0412308}.

\bibitem{Ratti:2005jh}
\bibinfo{author}{\bibfnamefont{C.}~\bibnamefont{Ratti}},
  \bibinfo{author}{\bibfnamefont{M.~A.} \bibnamefont{Thaler}},
  \bibnamefont{and} \bibinfo{author}{\bibfnamefont{W.}~\bibnamefont{Weise}},
  \bibinfo{journal}{Phys. Rev.} \textbf{\bibinfo{volume}{D73}},
  \bibinfo{pages}{014019} (\bibinfo{year}{2006}{\natexlab{b}}),
  \eprint{hep-ph/0506234}.

\bibitem{Ghosh:2006qh}
S.~K. Ghosh, T.~K. Mukherjee, M.~G. Mustafa, and R.~Ray, \emph{Phys. Rev.}
  \textbf{D73}, 114007 (2006), \eprint{hep-ph/0603050}.

\bibitem{Gava:1981qd}
\bibinfo{author}{\bibfnamefont{E.}~\bibnamefont{Gava}} \bibnamefont{and}
  \bibinfo{author}{\bibfnamefont{R.}~\bibnamefont{Jengo}},
  \bibinfo{journal}{Phys. Lett.} \textbf{\bibinfo{volume}{B105}},
  \bibinfo{pages}{285} (\bibinfo{year}{1981}).

\bibitem{Zantow:2003uh}
\bibinfo{author}{\bibfnamefont{F.}~\bibnamefont{Zantow}}
  (\bibinfo{year}{2003}), \eprint{hep-lat/0301014}.

\bibitem{Kaczmarek:2002mc}
\bibinfo{author}{\bibfnamefont{O.}~\bibnamefont{Kaczmarek}},
  \bibinfo{author}{\bibfnamefont{F.}~\bibnamefont{Karsch}},
  \bibinfo{author}{\bibfnamefont{P.}~\bibnamefont{Petreczky}},
  \bibnamefont{and} \bibinfo{author}{\bibfnamefont{F.}~\bibnamefont{Zantow}},
  \bibinfo{journal}{Phys. Lett.} \textbf{\bibinfo{volume}{B543}},
  \bibinfo{pages}{41} (\bibinfo{year}{2002}), \eprint{hep-lat/0207002}.

\bibitem{Kaczmarek:2005ui}
\bibinfo{author}{\bibfnamefont{O.}~\bibnamefont{Kaczmarek}} \bibnamefont{and}
  \bibinfo{author}{\bibfnamefont{F.}~\bibnamefont{Zantow}},
  \bibinfo{journal}{Phys. Rev.} \textbf{\bibinfo{volume}{D71}},
  \bibinfo{pages}{114510} (\bibinfo{year}{2005}), \eprint{hep-lat/0503017}.

\bibitem{Megias:2005ve}
\bibinfo{author}{\bibfnamefont{E.}~\bibnamefont{Meg{\'\i}as}},
  \bibinfo{author}{\bibfnamefont{E.}~\bibnamefont{Ruiz~Arriola}},
  \bibnamefont{and} \bibinfo{author}{\bibfnamefont{L.~L.}
  \bibnamefont{Salcedo}}, \bibinfo{journal}{JHEP}
  \textbf{\bibinfo{volume}{0601}}, \bibinfo{pages}{073} (\bibinfo{year}{2006}),
  \eprint{hep-ph/0505215}.

\bibitem{Dumitru:2003hp}
\bibinfo{author}{\bibfnamefont{A.}~\bibnamefont{Dumitru}}, 
\bibinfo{author}{\bibfnamefont{Y.}~\bibnamefont{Hatta}}, 
\bibinfo{author}{\bibfnamefont{J.}~\bibnamefont{Lenaghan}}, 
\bibinfo{author}{\bibfnamefont{K.}~\bibnamefont{Orginos}}, 
\bibnamefont{and}
\bibinfo{author}{\bibfnamefont{R.~D.}~\bibnamefont{Pisarski}},
\bibinfo{journal}{Phys. Rev.} \textbf{\bibinfo{volume}{D70}}, 
\bibinfo{pages}{034511} (\bibinfo{year}{2004}), \eprint{hep-th/0311223}.


\end{thebibliography}

\end{document}